\documentstyle[twoside,fleqn,espcrc2,epsf]{article}

\newcommand{\AmS}{{\protect\the\textfont2
  A\kern-.1667em\lower.5ex\hbox{M}\kern-.125emS}}

\title{ First study of $B \rightarrow \pi$ semileptonic decay 
        form factors using NRQCD  } 

\author{
H. Matsufuru\address{
   Department of Physics, Hiroshima University,
   Higashi-Hiroshima 739, Japan}\thanks{
   Presented by H. Matsufuru.  H.M. would like to thank
   the JSPS for Young Scientists for a research fellowship.},
S. Hashimoto\address{
   High Energy Accelerator Research Organization(KEK), 
   Tsukuba 305, Japan}\thanks{
   S.H. is supported by Ministry of Education, 
   Science and Culture under grant number 09740226.},
K-I. Ishikawa$^{\rm a}$, T. Onogi$^{\rm a}$, and N. Yamada$^{\rm a}$ }

\begin{document}

\begin{abstract}
We present a quenched calculation of the form factors of the semileptonic 
weak decay $B \rightarrow \pi l \bar{\nu}$ with $O(1/m_Q)$ NRQCD heavy
quark and Wilson light quark on a $16^3 \times 32$ lattice at $\beta=5.8$.
The form factors are evaluated at six heavy quark masses, 
in the range of $m_Q \sim 1.5-8$ GeV.
$1/m_Q$ dependence of matrix elements are investigated and 
compared with HQET predictions.
We observe clear signal for the form factors near $q^2_{\,max}$,
even at the $b$-quark mass range.
$f^0(q^2_{\,max})$ is compared with $f_B/f_{\pi}$ based on the 
soft pion theorem and significant difference is observed.

\end{abstract}

\maketitle

\section{Introduction}

Lattice study of $B$ decay matrix elements is important 
for the determination of 
Cabibbo-Kobayashi-Maskawa matrix elements,
and for investigations of applicability of 
Heavy quark effective theory ( HQET ) which is extensively
applied to phenomenological studies.
In this work, we calculate $B\rightarrow \pi$ form factors 
using the heavy quark described by $O(1/m_Q)$ NRQCD and Wilson 
light quark \cite{Onogi97a}.

\section{Simulation}

Numerical simulation is performed on 120 configurations of a 
$16^3 \times 32$ lattice at $\beta=5.8$ in the quenched approximation.
Wilson light quark is employed with three values of $\kappa$, 
0.1570, 0.1585 and 0.1600, which leads $\kappa_{c} = 0.163461(69)$
and lattice cutoff $a^{-1} = 1.714(63)$ GeV determined from
$\rho$ meson mass.
Light quark field is normalized with the factor 
$\sqrt{1-\frac{3 \kappa}{4 \kappa_c}}$ \cite{EKM97}.

The heavy quark is described 
with NRQCD including $O(1m_Q)$ corrections which leads following
evolution equation.
\[ G_{\varphi}(t\!=\!1)   
 \!=\!  \left( 1\!-\!\frac{1}{2n}H_0 \right)^n \!
     U_4^{\dag} \! \left( 1\!-\!\frac{1}{2n}H_0 \right)^n \! 
   G_{\varphi}(0), \] 
\[ G_{\varphi}(t\!+\!1)   
 \!=\!  \left( 1\!-\!\frac{1}{2n}H_0 \right)^n \!
     U_4^{\dag} \! \left( 1\!-\!\frac{1}{2n}H_0 \right)^n \] 
\begin{equation} 
\mbox{\hspace{4cm}} \times  ( 1 \!-\! \delta H ) \, G_{\varphi}(t).
\label{eq:evolve1}
\end{equation}
\begin{equation}
 H_0       =  -\frac{1}{2m_Q} \Delta^{(2)}, 
 \ \ \ \ \ 
 \delta H =  -\frac{1}{2m_Q} \: \vec{\sigma} \!\cdot\! \vec{B},
\end{equation}
where $\Delta^{(2)}$ denotes the lattice Laplacian and $B$ is
the chromomagnetic field.
The stabilizing parameter $n$ should satisfy $n>3/2m_Q$.

For the heavy quark, eight values of mass and stabilizing parameter
are used:  $(m_Q,n)=(5.0,1)$, $(2.6,1)$, $(2.1,1)$, $(2.1,2)$, $(1.5,2)$,
$(1.2,2)$, $(1.2,3)$, and $(0.9,2)$.
$m_{Q}=2.6$ and $0.9$ roughly correspond to the $b$- and $c$-quark 
masses.
The mean-field improvement\cite{LM93} is applied to the heavy quark 
evolution equation with 
$u_{0}=\langle\,\frac{1}{3}U_{plaq.}\,\rangle^{1/4}=0.867994(13)$.

The matrix elements are extracted from three point correlation functions, 
\[
C^{(3)}_{\;\mu}(p,k;t_{\!f},t_{\!s},t_{\!i}) 
 = \sum_{\vec{x}_{\!f}} \sum_{\vec{x}_{\!s}} \;
   e^{-i\vec{p} \,\cdot\, \vec{x}_{\!f}} \;
   e^{-i(\vec{k}-\vec{p}) \,\cdot\, \vec{x_{\!s}}} \;   \]
\vspace{-0.6cm} 
\[ \hspace*{2.6cm}  \times \langle  O_B( x_{\!f} ) 
        V^{ \dag}_{\mu}(x_{\!s} ) 
        O_{\pi}(t_{\!i},0 )  \rangle.
\]
We use 20 rotationally nonequivalent sets of $(\vec{p}, \pi)$
with $|\vec{p}|$, $|\vec{k}| \; \leq \sqrt{3}\cdot 2\pi/16$.
The source and the current operators are set on the time slices $t_i=4$
and $t_s=14$ respectively.
The matrix elements are extracted in the region $t_f=23-28$.

Numerical simulations were carried out on Intel Paragon XP/S
at INSAM (Institute for Numerical Simulation and Applied
Mathematics) in Hiroshima University.

We estimate the effect of perturbative corrections to the
heavy quark self-energy and the current \cite{Ishikawa97}.
In some cases, we use larger values of $n$ for the perturbation than
those in the simulation:
$(m_Q,n)=(5.0,1)$, $(2.6,2)$, $(2.1,2)$, $(1.5,3)$, $(1.2,3)$, and $(0.9,6)$.
This is because of the singularities encountered in the perturbative
expressions for some set of $(m_Q,n)$ with small $n$.
The multiplicative part of the current renormalization constant
is calculated with massless Wison quark for vanishing external momenta.
Two scales $q^*=\pi/a$ and $1/a$ are considered to define the expansion 
parameter $g_V^{\,2}$.

\section{Results}

\begin{figure}[tb]
\vspace*{-0.28cm}
\centerline{\epsfysize=5.4cm \epsffile{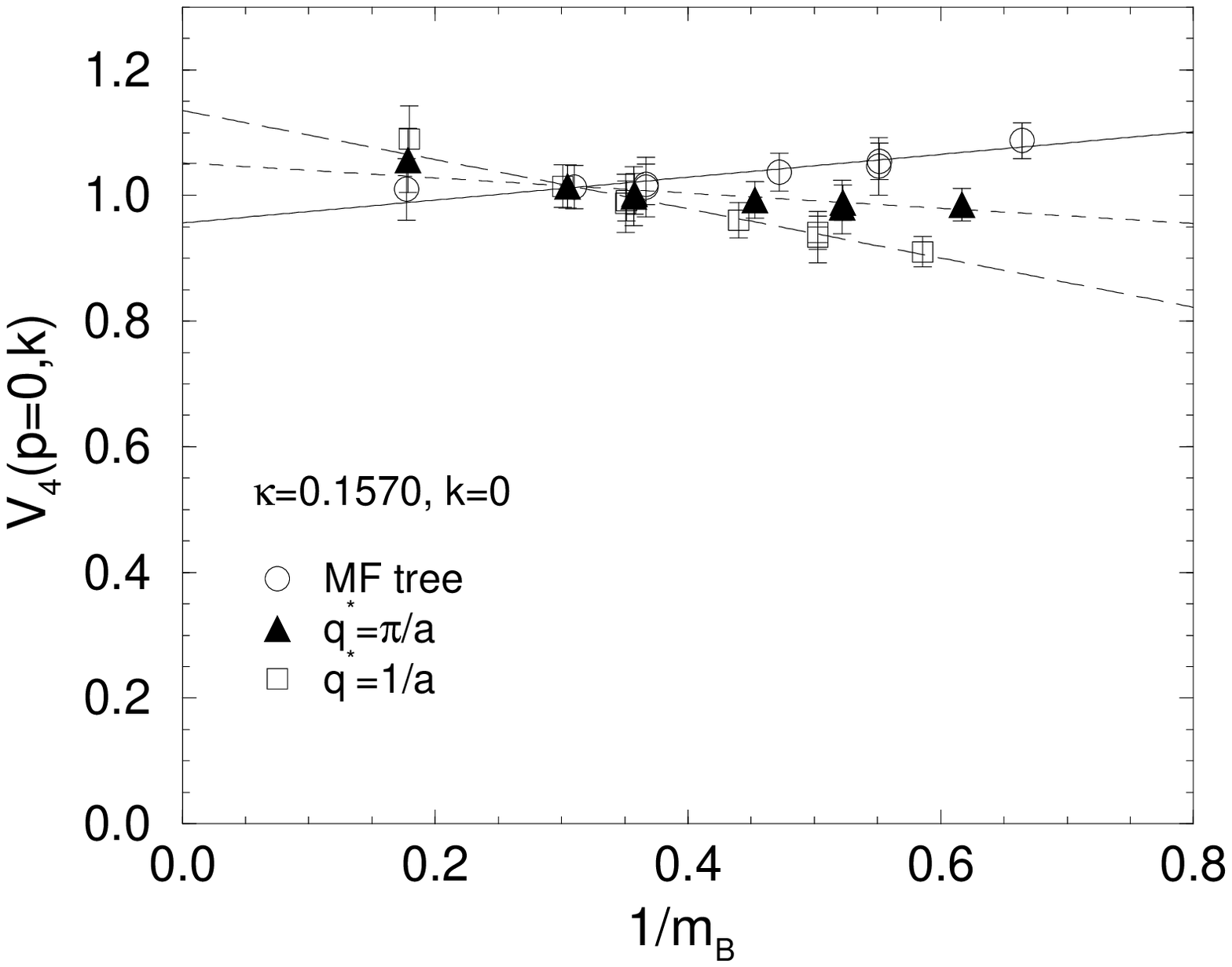}}
\vspace{-1.15cm}
\centerline{\epsfysize=5.4cm \epsffile{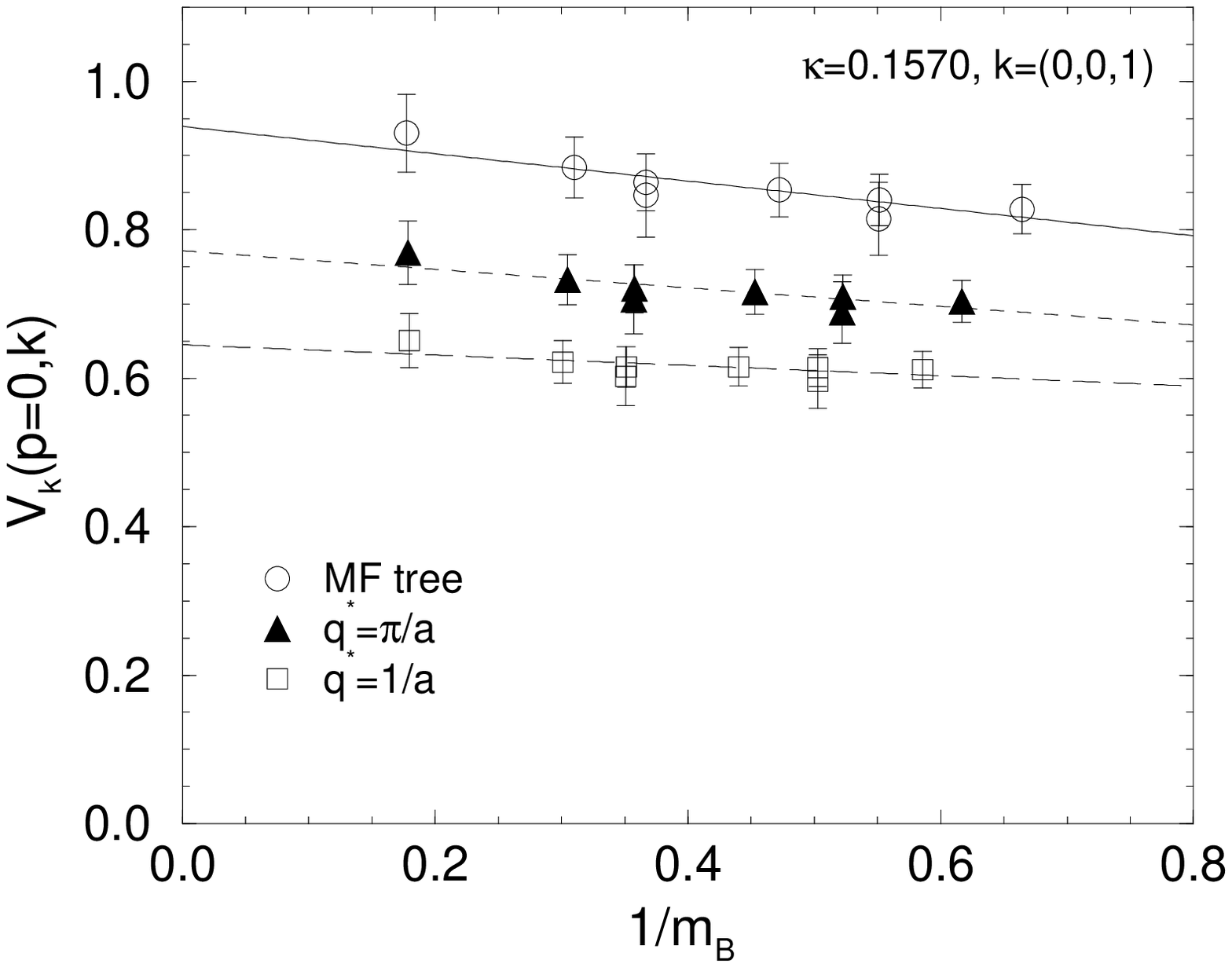}}
\vspace{-1.15cm}
\centerline{\epsfysize=5.4cm \epsffile{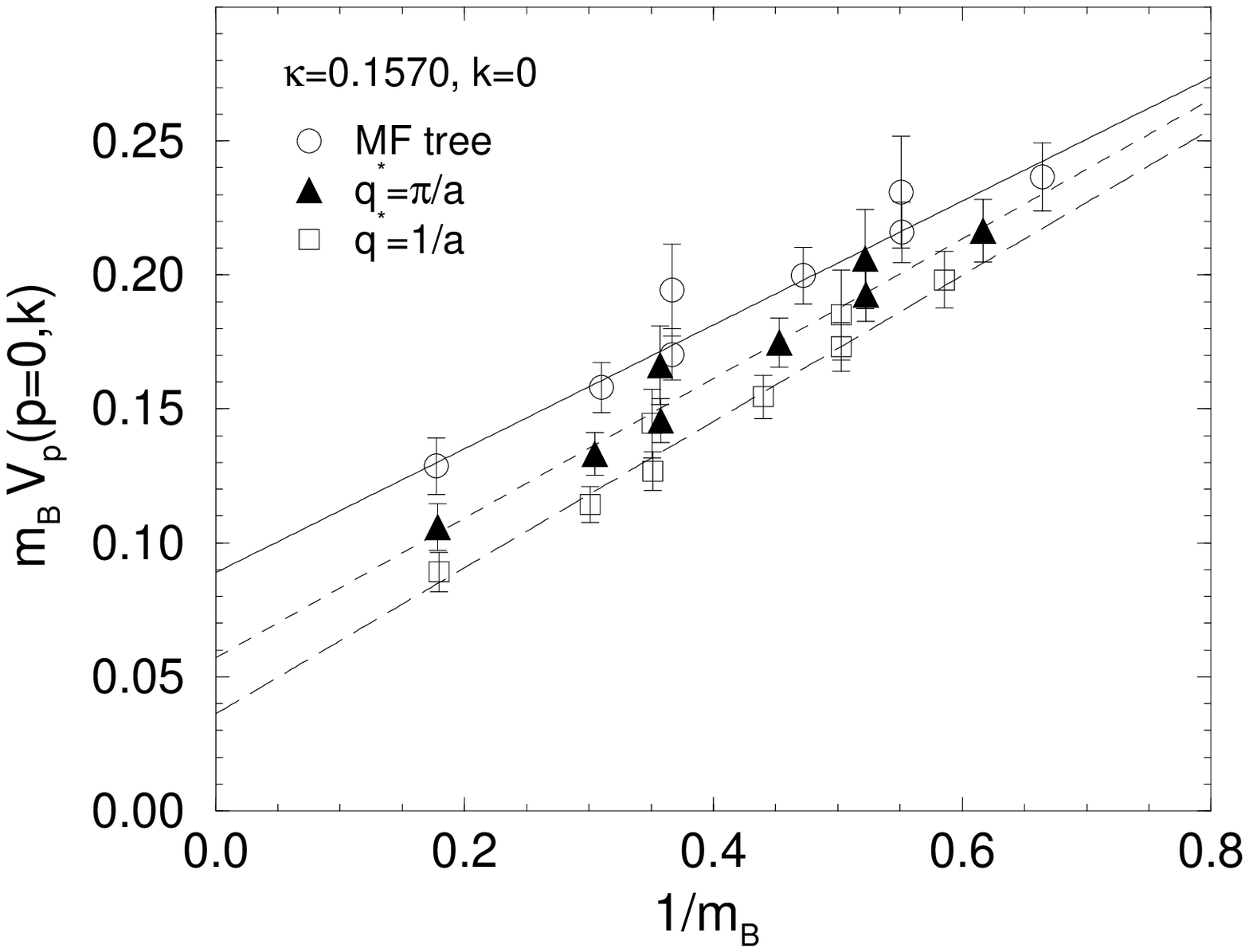}}
\vspace{-1.5cm}
\caption{$m_B$ dependence of $\hat{V}$ for $\kappa=0.1570$. 
$\hat{V}_4$ (top figure), $\hat{V}_k$ (middle), and $m_B \hat{V}_p$
(bottom) with three renormalization scales for each.}
\label{fig:ME}
\vspace{-0.2cm}
\end{figure}

It is useful to define following quantity.
\begin{equation}
   \hat{V}_{\mu}( \,\vec{p}, \vec{k}\, )  = 
  \frac{\langle \: \pi \,(\,\vec{k}\,) \: |\: V_{\mu} \: |\: 
    B\,(\, \vec{p}\, )\: \rangle }
    { \sqrt{\: 2\, E_{\pi}(k)}\; \sqrt{\: 2\, E_{B}(p)} }
\end{equation}
This expression can be entirely composed of numerical results, 
without any assumption  such as a dispersion relation.
It is also convenient for a comparison with HQET predictions.
According the heavy quark symmetry, 
for $\vec{p}=0$, $\hat{V}_{\mu}$ takes constant value
in the leading order of $1/m_Q$:
\begin{eqnarray}
 \hat{V}_4(\vec{p}=0,\vec{k}) 
  \!\!&=&\!\!  
   \hat{V}_4^{(0)}\left[ \, 1 + c_4^{(1)}/m_B + \cdots \right], \\
 \hat{V}_k(\vec{p}=0,\vec{k})
  \!\!&\equiv&\!\! \vec{\hat{k}}\cdot \vec{\hat{V}}(\vec{p}\!=\!0,\vec{k})
                                                    / \vec{\hat{k}}^2 
  \nonumber \\
  \!\!&=&\!\!  
  \hat{V}_k^{(0)} \left[ \, 1 + c_k^{(1)}/m_B + \cdots \right],
\end{eqnarray}
where $\hat{k}_i=2\sin(k_i/2)$.
Upper two of Figure \ref{fig:ME} show the results of
$\hat{V}_4$ for $\vec{p}\!\!=\!\!\vec{k}\!\!=\!\!0$ and 
$\hat{V}_k$ for $|\vec{k}|=2\pi/16$ in the case of $\kappa=0.1570$. 
They are evaluated at three renormalization scales, 
mean-field tree, $q^*=\pi/a$ and $1/a$.
Both $\hat{V}_4$ and $\hat{V}_k$ less depend on $m_B$
in comparison with $f_B$ case \cite{Yamada97}.
The spacial component of $\hat{V}$ is more affected by the perturbative 
corrections than the temporal one is.
It is also predicted that
\begin{eqnarray}
 \hat{V}_p(\vec{p}=0,\vec{k}) 
 &\equiv&  \mathop{\mbox{lim}}_{\vec{p}^{\,2}\rightarrow 0}  
         \vec{\hat{p}}\cdot \vec{\hat{V}}(\vec{p},\vec{k})
                                        / \; \vec{\hat{p}}^{\,2}  
 \nonumber \\
 & &\hspace{-1.5cm} = \frac{1}{m_B} \hat{V}_p^{\prime (0)} \left[ \, 1 + 
                             c_p^{(1)}/m_B + \cdots \right].
\end{eqnarray}
We extrapolate $\hat{V}_p$ at finite $\vec{p}$ to 
$\vec{p}\!=\!\!0$
linearly in $\vec{p}^{\,2}$ to determine 
$\hat{V}_p(\vec{p}\!\!=\!\!0,\vec{k})$.
$\hat{V}_p(\vec{p}\!=\!0,\vec{k}\!=\!0)$ multiplied by $m_B$ 
is also displayed in Figure \ref{fig:ME}. 
Contrary to the cases of $\hat{V}_4$ and $\hat{V}_k$, 
$O(1/m_B)$ effect is significant for $\hat{V}_p$.

\begin{figure}[tbh]
\vspace*{-0.28cm}
\centerline{\epsfysize=6.2cm \epsffile{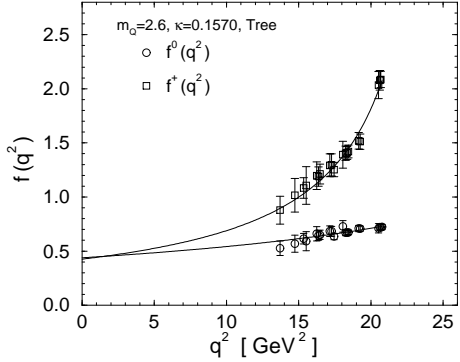}}
\vspace{-1.5cm}
\caption{Form factors for $m_Q=2.6$ and $\kappa=0.1570$ in the mean-field 
tree level. Solid lines correspond to the results of single pole fit. } 
\label{fig:FF}
\vspace{-0.2cm}
\end{figure}

The matrix elements are expressed in terms of two form factors, 
$f^0$ and $f^+$ :
\begin{eqnarray}
\langle \pi(k) | V_{\mu} | B(p) \rangle \!\!\!\! &=& \!\!\!\! 
 \left( p + k - q \frac{ m_B^{\,2} \!-\! m_{\pi}^{\,2} }{q^2} \right)_{\!\mu}
 f^{+}(q^2)   \nonumber \\
 & & + q_{\mu}\frac{ m_B^{\,2}  - m_{\pi}^{\,2} }{q^2} f^{0}(q^2).
\end{eqnarray}
To obtain the form factors from $\hat{V}_{\mu}$, it is necessary 
to assume some dispersion relations for $\pi$ and $B$.
For both of $B$ and $\pi$, we employ the form 
$E(p)^2=m^2+\sum_{i}4\sin^2 (p_i/2)$ to assure the on-shell conditions.
Figure \ref{fig:FF} shows the obtained form factors for $m_Q=2.6$, 
$\kappa=0.1570$, in the mean-field tree level.
It is observed that for larger $m_Q$, $f^+$ increase rapidly near 
$q^2_{\,max}$, which may be explained as the $B^*$ pole
contribution.
We fit the form factors to a single pole function, whose result
is displayed in Figure \ref{fig:FF} as the solid lines.
The fit of $f^+$ gives the exchanged vector particle mass
$m_{pole}=2.979(49)$ which is comparable with $m_{B^*}=3.2502(62)$ 
determined from the two point correlator.
The renormalization corrections change the shape of $f^+$ more than
$f^0$, because of asymmetric effect on temporal and spacial components.
Even after the renormalization is incorporated, the pole fit gives
similar $m_{pole}$.

We extrapolate the matrix elements linearly in $1/\kappa$ 
to the chiral limit.
Alternative extrapolation using the form factors gives
7 \% discrepancy of $f^0(q^2_{\,max})$ 
for $m_Q=2.6$ and 14 \% for $m_Q=0.9$.
At the zero recoil point, the prediction of heavy meson effective theory
\cite{KK94} 
advocates the linear extrapolation of the matrix element.
The results at $\kappa_c$ are much noisy for quantitative discussions. 

\begin{figure}[t]
\vspace*{-0.28cm}
\centerline{\epsfysize=6.2cm \epsffile{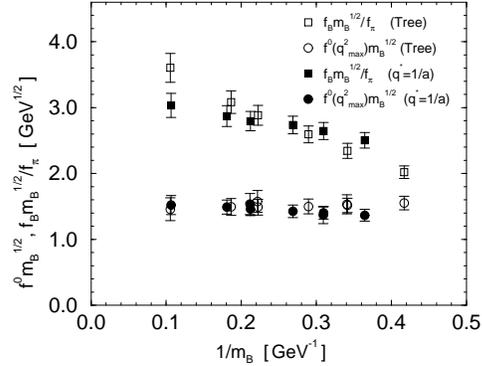}}
\vspace{-1.5cm}
\caption{Comparison of $f^0(q^2_{\,max})$ with $f_B/f_{\pi}$ 
in two scales, mean-field tree and $q^*=1/a$. }
\label{fig:SPT}
\vspace{-0.2cm}
\end{figure}

Finally, we consider the implication of soft pion theorem 
\cite{BD92,KK94}.
For the massless pion limit, $f^0(q^2_{\, max})$ should
equal to $f_B/f_{\pi}$.
This relation is examined in Figure \ref{fig:SPT},
using our result on $f_B$ determined with slightly different 
form of NRQCD \cite{Yamada97}. 
Significant difference is observed in large $m_B$ region.
Similar result is obtained in the work using Fermilab action for 
the heavy quark \cite{Tominaga97}.
The origin and physical meaning of this discrepancy remains 
as a future problem.

\end{document}